\documentclass{aastex702}

\usepackage{amsmath}

\shorttitle{Age-Resolved Warp Kinematics}
\shortauthors{Zhou et al.}
\graphicspath{{./}}

\newcommand{\kms}{km s$^{-1}$}

\begin{document}

\title{Age-Resolved Vertical Motions in the Milky Way Warp: A Localized Enhancement in Intermediate-Age Populations}

\correspondingauthor{Xiaoyue Zhou, Xiaodian Chen, and Haifeng Wang}
\email{zhouxy@bao.ac.cn; chenxiaodian@nao.cas.cn; haifeng.wang.astro@gmail.com}

\author[0009-0001-5141-9130]{Xiaoyue Zhou}
\email{zhouxy@bao.ac.cn}
\affiliation{CAS Key Laboratory of Optical Astronomy, National Astronomical Observatories, Chinese Academy of Sciences, Beijing 100101, China}
\affiliation{School of Astronomy and Space Science, University of the Chinese Academy of Sciences, Beijing, 100049, China}

\author[0000-0001-7084-0484]{Xiaodian Chen}
\email{chenxiaodian@nao.cas.cn}
\affiliation{CAS Key Laboratory of Optical Astronomy, National Astronomical Observatories, Chinese Academy of Sciences, Beijing 100101, China}
\affiliation{School of Astronomy and Space Science, University of the Chinese Academy of Sciences, Beijing, 100049, China}
\affiliation{Institute for Frontiers in Astronomy and Astrophysics, Beijing Normal University, Beijing 102206, China}

\author[0000-0001-8459-1036]{Haifeng Wang}
\email{haifeng.wang.astro@gmail.com}
\affiliation{Department of Astronomy, China West Normal University, Nanchong, 637009, China}
\affiliation{Local Universe and Time-Domain Astronomy Laboratory, LuoxiaHong Institute of Astronomy, China West Normal University, Nanchong, 637002, China}

\author[0000-0003-4489-9794]{Shu Wang}
\email{}
\affiliation{CAS Key Laboratory of Optical Astronomy, National Astronomical Observatories, Chinese Academy of Sciences, Beijing 100101, China}
\affiliation{Department of Astronomy, China West Normal University, Nanchong, 637009, China}

\author[0000-0002-6989-8192]{Zhihong He}
\email{}
\affiliation{Department of Astronomy, China West Normal University, Nanchong, 637009, China}

\author[0000-0001-9073-9914]{Licai Deng}
\email{}
\affiliation{CAS Key Laboratory of Optical Astronomy, National Astronomical Observatories, Chinese Academy of Sciences, Beijing 100101, China}
\affiliation{School of Astronomy and Space Science, University of the Chinese Academy of Sciences, Beijing, 100049, China}
\affiliation{Department of Astronomy, China West Normal University, Nanchong, 637009, China}

\begin{abstract}
Determining how the Milky Way warp evolves provides constraints on the formation and dynamical evolution of galactic disks.
Using Apache Point Observatory Galactic Evolution Experiment Data Release 19 (APOGEE DR19) spectroscopy, we estimate ages for 222,355 red giants. With Gaia astrometry and geometric distances, we analyze 168,824 of these stars in five age bins spanning $0$--$10$ Gyr. The warp-related radial vertical-velocity ($V_Z$) profile of young red giants ($0$--$2$ Gyr) is broadly consistent with those of classical Cepheids and young open clusters. Over $9\leq R<12$ kpc, we find that the mean $V_Z$ values of the intermediate-age red giants ($2$--$4$ and $4$--$6$ Gyr) are $1.48\pm0.46$ and $1.49\pm0.52$ \kms\ higher than that of the young population, respectively. By contrast, at $R\simeq12$--$13$ kpc, the $V_Z$ profiles of the $0$--$6$ Gyr red giants are consistent within their uncertainties. The intermediate-age enhancement persists across our robustness tests and may reflect smaller effective kinematic warp onset radii for the intermediate-age populations. The observed velocity differences may arise from the combined effects of earlier perturbations and the present-day warp, or from differences in vertical structure among stellar populations.
\end{abstract}


\section{Introduction} \label{sec:intro}
The Milky Way disk warp is one of the most prominent large-scale three-dimensional structures in the outer disk, in which the disk plane bends systematically away from the Galactic midplane with radius and azimuth \citep{2019chen}. The warp is traced not only by the spatial distributions of stars and gas, but also by ordered signatures in the mean vertical velocity ($V_Z$) \citep{2018poggio,zhou2024}. Thus, $V_Z$ complements geometric measurements of outer-disk dynamics \citep{poggio2020,zhou2025}. The origin of the warp may involve tidal perturbations by satellite galaxies \citep{2006ApJ...641L..33W}, misalignment between the dark matter halo and the Galactic disk \citep{2023NatAs...7.1481H}, and intergalactic gas accretion \citep{2002A&A...386..169L}. The warp therefore provides a probe of perturbations acting on the outer disk and of the dynamical evolution of the Milky Way.

Previous studies have measured the Galactic warp using a variety of tracers, including neutral atomic hydrogen (H~I) gas \citep{1957AJ.....62...90B,1957Natur.180..677K}, CO molecular clouds \citep{sun2026co}, O- and B-type stars \citep{2019A&A...627A.150R}, classical Cepheids (CCs) \citep{2019chen, 2019Sci...365..478S,zhou2025}, red clump (RC) stars \citep{2024MNRAS.527.4863U}, red giants \citep{2019A&A...627A.150R,2020ApJ...905...49C}, and open clusters (OCs) \citep{2023ApJ...954L...9H,2024arXiv241220344P}. Although these studies consistently reveal the outer-disk warp, the inferred warp amplitude, onset radius, orientation of the line of nodes (LON), and $V_Z$ pattern vary among tracers \citep{2019A&A...627A.150R,cabreragadea2024}. These differences may reflect real variations in the warp signatures traced by different stellar populations, or they may be affected by the spatial coverage, selection effects, distance uncertainties, and model assumptions. Interpreting these differences is therefore essential for using the Galactic warp as a dynamical tracer.

Some studies report that warp kinematic signals weaken with stellar age \citep{2017IAUS..321....6L,wang2020,li2023}. Recent mono-age RC mapping likewise reports population-dependent variation in the vertical warp signal \citep{sun2024}; meanwhile, other work suggests that older or intermediate-age populations may exhibit larger geometric warp amplitudes, with substantial scatter among existing measurements \citep{2024MNRAS.527.4863U,cabreragadea2024,kushwahaa2026}. Thus, no unified picture of the age dependence of the Galactic warp has yet emerged, motivating a systematic multi-tracer comparison of outer-disk vertical kinematics using age-resolved red giants and complementary young tracers.

Red giants are luminous and abundant, span a broad age range, and have ages that can be estimated from their spectra, enabling age-resolved studies of outer-disk structure and kinematics \citep{2016MNRAS.456.3655M,2023anders}. The Apache Point Observatory Galactic Evolution Experiment (APOGEE) provides high-resolution near-infrared spectra, precise radial velocities, and detailed chemical abundances \citep{2017AJ....154...94M,2025AJ....170...96M}; its combination with Gaia astrometry yields six-dimensional phase-space information \citep{2023A&A...674A...1G}. Compared with APOGEE Data Release 17 (DR17), APOGEE DR19 provides approximately 1.2 million near-infrared spectral products \citep{sdssdr19}, while our red-giant age catalog is approximately 24\% larger than that of \citet{2023anders}.

In this paper, we directly compare the observed radial profiles of $V_Z$ for
APOGEE--Gaia red giants spanning 0--10 Gyr, CCs, and OCs. Without imposing a parametric warp model, we test whether the outer-disk $V_Z$ signal weakens monotonically with age and is consistent among young tracers.

\section{Data and Methodology} \label{sec:data}
We construct an age-resolved APOGEE DR19 red-giant sample using spectroscopic ages, APOGEE radial velocities, Gaia astrometry, and the Bailer-Jones median geometric distances, $r_{\rm med,geo}$ \citep[hereafter BJ21]{2021AJ....161..147B}. We analyze the parent sample and its non-RC and RC subsamples, supplemented by CCs and OCs as young tracers. All tracers are transformed to a common Galactocentric reference frame. Details of the spectroscopic-age inference and the alternative RC photometric-distance test are provided in Appendices~\ref{app:ages} and~\ref{app:distance}, respectively.

\subsection{Spectroscopic Ages of APOGEE DR19 Red Giants} \label{subsec:ages}
We use APOGEE DR19 stellar parameters and elemental abundances from the APOGEE Stellar Parameters and Chemical Abundances Pipeline (ASPCAP), provided by the Astra framework of the Sloan Digital Sky Survey V (SDSS-V) \citep{sdssdr19,2025AJ....170...96M}. First dredge-up induces mass-dependent changes in the surface carbon and nitrogen abundances of red giants, allowing C/N-sensitive abundance information to provide empirical constraints on stellar mass and age. We therefore estimate stellar ages using a multivariate spectroscopic model calibrated on red giants with asteroseismic ages \citep{2015MNRAS.453.1855M,2016MNRAS.456.3655M,ness2016}.

Following \citet{2023anders}, we retrain an XGBoost spectroscopic-age model for APOGEE DR19 red giants using APOGEE--Kepler asteroseismic ages from \citet{miglio2021} as calibration labels. After quality selection, the calibration sample contains 2,598 stars. The model uses effective temperature $T_{\rm eff}$, surface gravity $\log g$, and 15 abundance ratios as input features, and achieves a root-mean-square error (RMSE) of 1.20 Gyr on the held-out test set, with a coefficient of determination $R^2=0.848$ and a mean residual of 0.05 Gyr (see Figure~\ref{fig:age_validation}).

We apply the model to APOGEE DR19 red giants within the training-sample parameter domain, requiring reliable spectra and complete input features. This yields a parent catalog of 222,355 red giants with spectroscopic ages, of which 218,739 have unflagged age estimates. The parent catalog has a median age uncertainty of 1.11 Gyr and a median relative age uncertainty of $20.4\%$. The sample selection, model training, age correction, uncertainty estimation, and validation are described in Appendix~\ref{app:ages}.

\subsection{Red-giant Kinematic Sample}
\label{subsec:rg_sample}

We construct the red-giant kinematic sample from the APOGEE DR19 red-giant age catalog described in Section~\ref{subsec:ages}. We combine APOGEE line-of-sight velocities ($V_{\rm los}$), Gaia DR3 proper motions, and BJ21 geometric distances to derive Galactocentric positions and velocities. We require valid distances and proper motions, unflagged spectroscopic ages, and APOGEE $V_{\rm los}$ with
\texttt{doppler\_flags=0}. These criteria retain 218,140 of the 222,355 red giants in the parent catalog. 

Following the coordinate conventions of \citet{zhou2024}, we transform the observables into Galactocentric cylindrical coordinates $(R,\phi,Z,V_R,V_\phi,V_Z)$. We place the Sun at $(X,Y,Z)=(R_0,0,0)$ with $R_0=8.277$ kpc and $Z_\odot=0$, and adopt a solar Galactocentric velocity of $(U_\odot,V_{\phi,\odot},W_\odot)=(9.3,251.5,8.59)$ \kms. Here $U_\odot$ points toward the Galactic center, $V_{\phi,\odot}$ follows Galactic rotation, and $W_\odot$ points toward the north Galactic pole. The azimuth $\phi$ is measured from the Galactic center--Sun line and increases with Galactic rotation; $V_R$ is positive outward, $V_\phi$ follows Galactic rotation, and $V_Z$ is positive toward the north Galactic pole.

For the $V_Z$ analysis, we restrict the sample to
$5<R<16$ kpc, $-60^\circ<\phi<60^\circ$, and $|Z|<1.5$ kpc. We remove extreme kinematic
outliers by requiring $|V_R|<75$ \kms, $|V_Z|<50$ \kms, and
$100<V_\phi<300$ \kms. The resulting 177,030 red giants define the
kinematic sample used in the subsequent analysis.

\begin{figure}
\centering
\includegraphics[width=\linewidth]{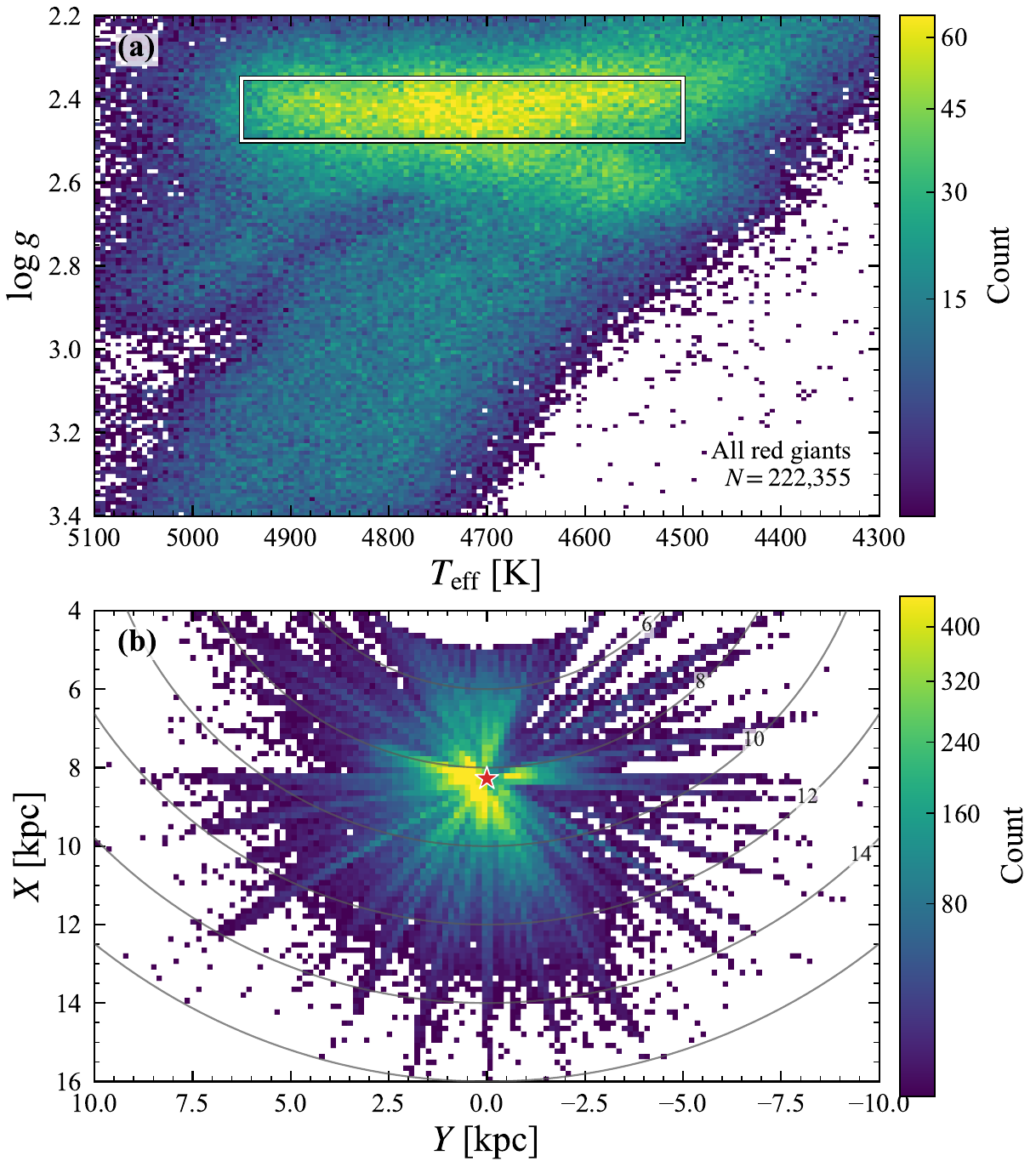}
\caption{APOGEE DR19 red-giant sample used in this work. Panel (a) shows the density distribution of the parent red-giant sample in the
$T_{\rm eff}$--$\log g$ plane; the white box marks the adopted RC selection. Panel (b) shows the Galactocentric $X$--$Y$ distribution of the final
red-giant kinematic sample using BJ21 geometric distances. Gray arcs mark Galactocentric radii, and the red star marks the Sun.}
\label{fig:rg_sample}
\end{figure}

To test whether the measured age dependence is driven by the RC subset,
we identify RC candidates using $4500\leq T_{\rm eff}\leq4950$ K and
$2.35\leq\log g\leq2.50$, following criteria similar to those of
\citet{2019ApJ...877..116W,2021ApJ...923..145W}. This selection divides
the kinematic sample into 50,193 RC stars and 126,837 non-RC red giants. We repeat the primary analysis using the non-RC subset. Figure~\ref{fig:rg_sample} summarizes the RC selection in the $T_{\rm eff}$--$\log g$ plane and the spatial coverage of the final red-giant kinematic sample.

\subsection{Age Binning and Vertical-Velocity Fields} \label{subsec:vz_fields}

For the age-resolved analysis, we divide the red-giant kinematic sample into five spectroscopic-age bins: $0$--$2$, $2$--$4$, $4$--$6$, $6$--$8$, and $8$--$10$ Gyr. These bins contain 13,081, 36,584, 53,783, 37,004, and 28,372 stars, respectively, for a total of 168,824 stars. The remaining 8,206 stars have spectroscopic ages $\geq10$ Gyr and are excluded owing to their lower age precision and valid $R$--$\phi$ coverage extending only to $R\simeq10.7$ kpc.

Since the red-giant sample has non-uniform coverage in the $R$--$\phi$ plane, we bin each age subsample using $\Delta R=0.5$ kpc and $\Delta\phi=5^\circ$, retaining cells with at least 20 stars. Each valid cell is represented by its median $R$, $\phi$, and $V_Z$, with the $V_Z$ uncertainty estimated from 1000 bootstrap resamples. 

\begin{figure*}
\centering
\includegraphics[width=\textwidth]{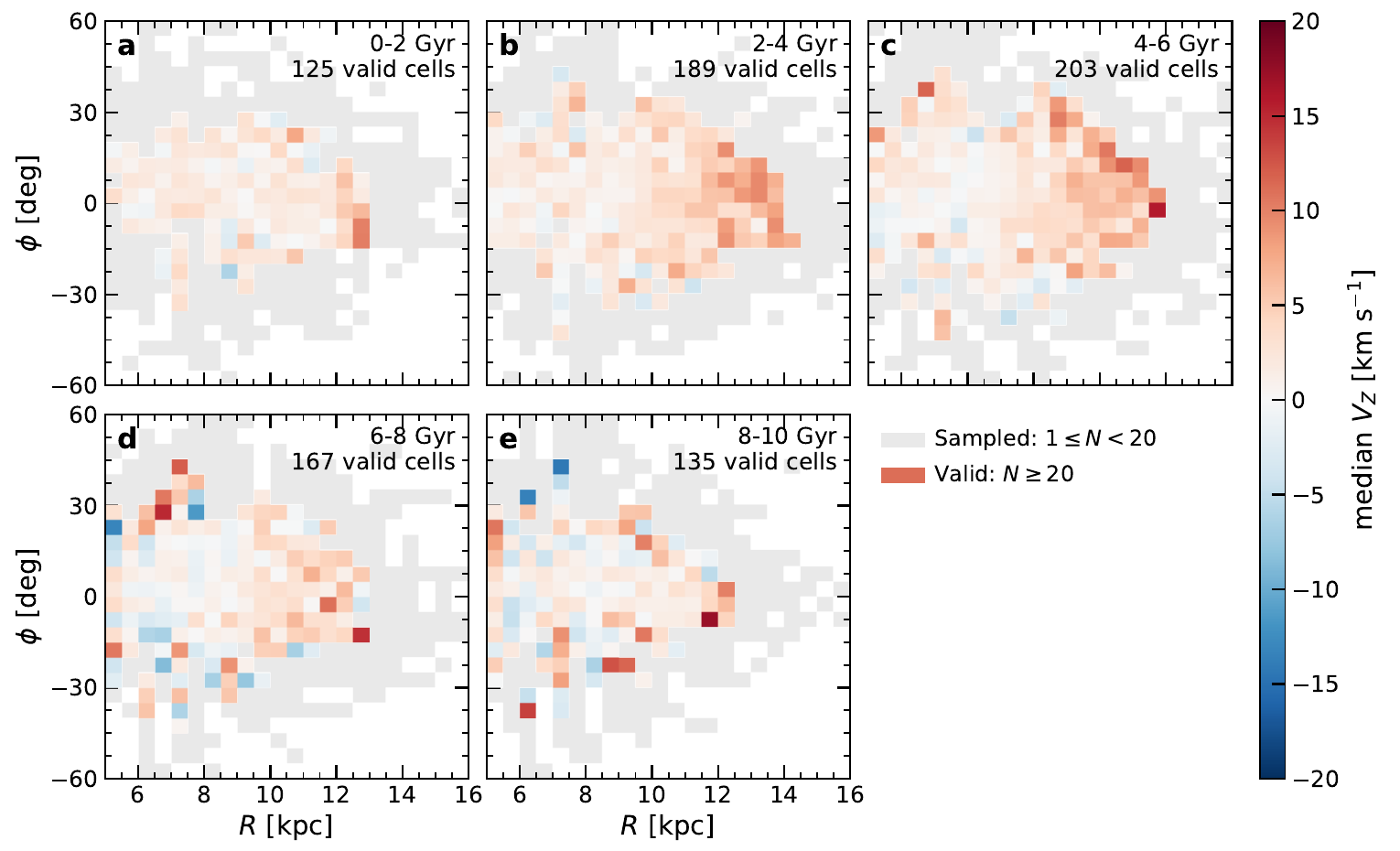}
\caption{Median vertical-velocity ($V_Z$) fields of the red-giant sample in five age bins. Colored cells show the median $V_Z$ in valid $R$--$\phi$ cells containing at least 20 stars; red and blue indicate positive and negative $V_Z$, respectively. Gray cells contain $1\leq N<20$ stars and are excluded from subsequent analysis, whereas white regions are unsampled. Panels are ordered by increasing age, with the number of valid cells given in each panel.}
\label{fig:spatial_bins}
\end{figure*}

Figure~\ref{fig:spatial_bins} shows the resulting median $V_Z$ fields. The $0$--$8$ Gyr populations all show positive $V_Z$ in the outer disk, indicating upward vertical motion. The signal is most spatially coherent in the $2$--$4$ and $4$--$6$ Gyr bins. The $6$--$8$ Gyr field retains a coherent positive $V_Z$ structure over $R\simeq9$--$13$ kpc and $|\phi|\lesssim20^\circ$, although the signal is weaker and less extended than in the $2$--$6$ Gyr bins. The $8$--$10$ Gyr field is less coherent and has the most limited outer-disk coverage. The positive outer-disk structures are
qualitatively consistent with warp-related vertical motions. Overall, the valid cells are concentrated at $R\simeq6$--$13$ kpc and $|\phi|\lesssim30^\circ$. Although this footprint does not sample all warp azimuths, it covers the sector where the positive warp-related $V_Z$ signal is strongest and supports the present age-resolved comparison. We assess the effect of spatial coverage using a common-footprint test in Section~\ref{subsec:robustness}.

\subsection{Young Reference Tracers}
\label{subsec:young_tracers}

We draw the young reference samples from 1,939 CCs in \citet{zhou2024} and the public catalog of 5,340 OCs in \citet{2023ApJ...954L...9H}. The CC distances were derived from the period--Wesenheit--metallicity relation, and their velocities were calculated using Gaia DR3 astrometry and mean line-of-sight velocities. For OCs, we apply a parallax zero-point correction of $+0.017$ mas to the cataloged mean Gaia DR3 parallaxes \citep{2021lindegren}. We require ages below 1 Gyr, $|V_{\rm los}|\leq50$ \kms, and at least one member with a measured $V_{\rm los}$. Both tracers are transformed using the same coordinate convention and solar motion adopted for the red giants.

We restrict both samples to $5<R<16$ kpc and $|\phi|<20^\circ$ and apply the same $Z$, $V_R$, $V_Z$, and $V_\phi$ cuts as for the red giants. We find that varying the azimuthal limit between $10^\circ$ and $40^\circ$ does not alter the qualitative $V_Z(R)$ trends. After these selections, the final samples contain 747 CCs and 2,186 OCs.

\section{Results and Discussion}
\label{sec:results}

\subsection{Age-dependent Vertical-velocity Profiles}
\label{subsec:age_profiles}

\begin{figure*}
\centering
\includegraphics[width=\textwidth]{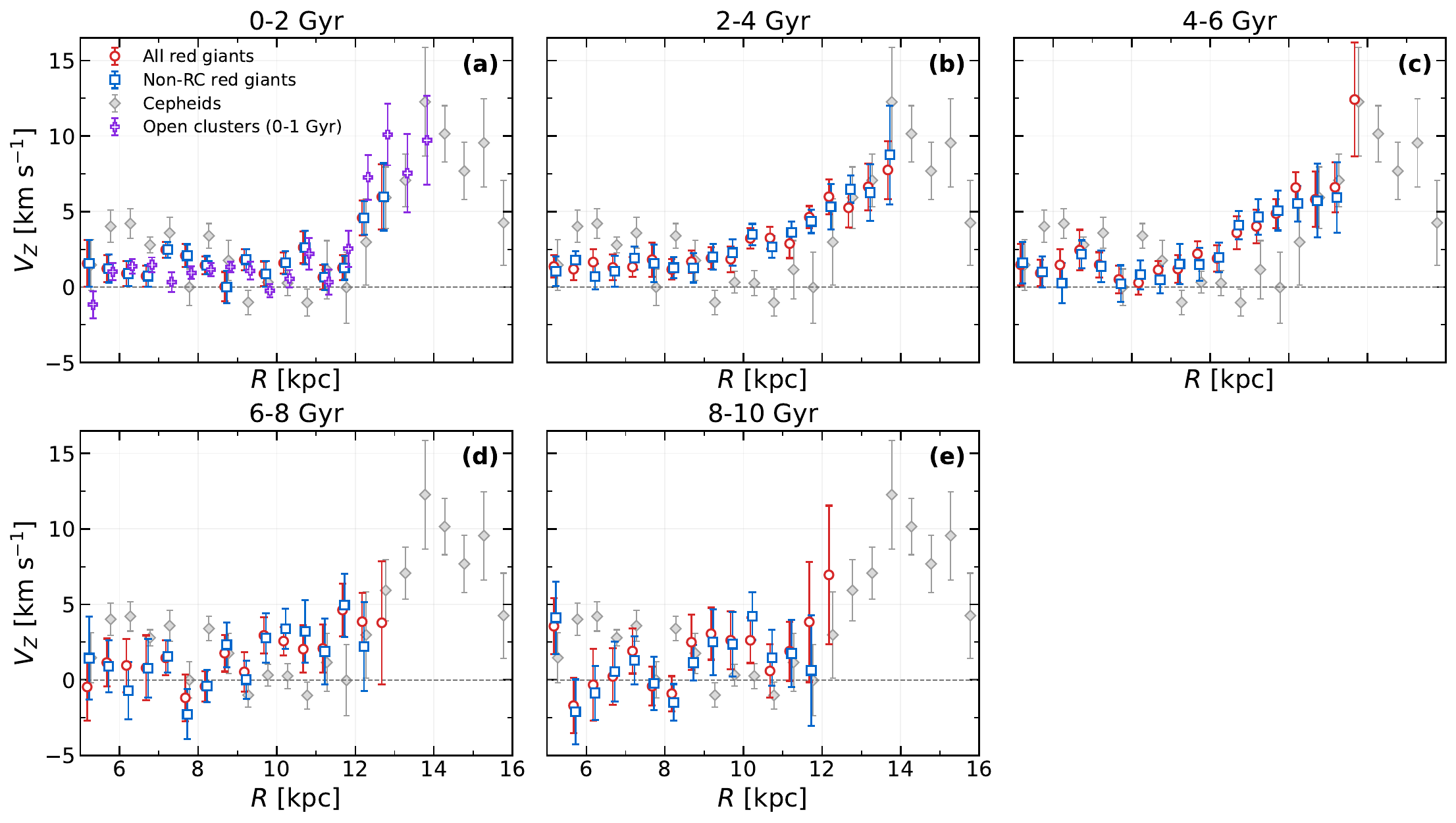}
\caption{Age-resolved radial profiles of the observed $V_Z$. Red circles show the full red-giant sample, while blue squares show its non-RC subset. Gray diamonds show the same Cepheid profile repeated in all panels as a common young reference. Purple symbols show open clusters younger than 1 Gyr and are included only in the $0$--$2$ Gyr panel. Error bars indicate $1\sigma$ uncertainties. In panel~(a), the red circles are mostly obscured by the blue squares.}
\label{fig:multitracer_vz}
\end{figure*}

We construct profiles of vertical velocity $V_Z$ as a function of Galactocentric radius $R$ for the full red-giant sample and its non-RC subset in five age bins spanning $0$--$10$ Gyr, and compare them with those of the young reference tracers, CCs and OCs (Figure~\ref{fig:multitracer_vz}).

For both red-giant samples, we calculate $V_Z$ in each age and radial bin as the mean of the median $V_Z$ values in valid spatial cells. We estimate the uncertainty using 1000 bootstrap--Monte Carlo realizations. In each realization, we resample the cells with replacement, keeping the number of cells unchanged, and perturb their median velocities according to their uncertainties. A minimum cell uncertainty of 1 \kms\ is adopted. We then calculate a mean velocity for each realization and adopt the standard deviation of these 1000 means as the $1\sigma$ uncertainty. Given the anticenter-dominated coverage (Figure~\ref{fig:spatial_bins}), we restrict the radial-profile analysis to $|\phi|<40^\circ$ and retain radial bins containing at least two valid spatial cells.

For CCs and OCs, we bin the samples directly in radius using 0.5 kpc bins. We require at least five CCs or three OCs per bin, yielding 22 CC bins and 18 OC bins. We adopt the median $V_Z$ in each bin and estimate its $1\sigma$ uncertainty using 1000 bootstrap resamples. The same CC profile is shown in every panel of Figure~\ref{fig:multitracer_vz} as a common reference, while OCs are shown only in the $0$--$2$ Gyr panel.

Figure~\ref{fig:multitracer_vz} shows similar $V_Z$ profiles for the full red-giant sample and its non-RC subset in each age bin. In panel~(a), the young red giants ($0$--$2$ Gyr), CCs, and OCs show broadly consistent radial $V_Z$ profiles. All three tracers show increasing $V_Z$ with $R$ over $R\simeq11$--$13$ kpc. The intermediate-age populations ($2$--$4$ and $4$--$6$ Gyr) in panels~(b) and~(c) also show increasing $V_Z$ with $R$ over $R\simeq9$--$14$ kpc. The main difference between the intermediate-age and young red giants occurs at $9\leq R<12$ kpc. In this radial range, the $2$--$4$ and $4$--$6$ Gyr populations have mean $V_Z$ values higher than that of the $0$--$2$ Gyr population by $1.48\pm0.46$ and $1.49\pm0.52$ \kms, respectively. These offsets are averages of the differences across six radial bins, with errors propagated from the bin uncertainties. At $R\simeq12$--$13$ kpc, the $0$--$6$ Gyr red-giant profiles are consistent within their uncertainties. The $6$--$8$ and $8$--$10$ Gyr populations in panels~(d) and~(e) show predominantly positive $V_Z$ over $R\simeq9$--$13$ and $9$--$12$ kpc, respectively. However, their larger velocity uncertainties and limited coverage at $R\gtrsim12$ kpc make it difficult to assess the warp-related kinematic signals in these older populations ($6$--$10$ Gyr).

The consistency of the radial profiles traced by the three young populations suggests that these stars are responding to the same large-scale warp structure. This agreement is notable because the three young tracers have independent distance scales, selection functions, and age estimates. It therefore provides a direct kinematic cross-check that the youngest APOGEE red giants trace the same outer-disk warp signal as the established young tracers in the surveyed radial and azimuthal range.

The higher $V_Z$ of intermediate-age populations relative to young populations at $R\simeq9$--$12$ kpc is qualitatively consistent with the population differences reported by \citet{2020ApJ...905...49C}, \citet{2021A&A...649A...8G}, and \citet{2025A&A...701A.136C}. \citet{2020ApJ...905...49C} found that both the $V_Z$ amplitude and the radius of its peak vary with stellar age. \citet{2021A&A...649A...8G} found oscillations in the radial $V_Z$ profiles, with a deeper dip near $R\simeq11$ kpc in young populations. Similarly, Figure~3 of \citet{2025A&A...701A.136C} shows higher mean $V_Z$ for thin-disk RR Lyrae stars than for classical Cepheids over $R\simeq10$--$12$ kpc. Their $V_Z$ dispersion suggests a kinematic age of approximately $3$--$4$ Gyr for the thin-disk RR Lyrae population. This trend is qualitatively consistent with our finding that $2$--$6$ Gyr red giants exhibit higher $V_Z$ than younger populations over a comparable radial range. By independently estimating spectroscopic ages, we further quantify this localized enhancement across distinct age populations over $9\leq R<12$ kpc.

In contrast to previous studies that inferred a decreasing warp amplitude with increasing stellar age from kinematic modeling \citep{wang2020,li2023}, we directly compare the observed $V_Z$ profiles of different age populations at fixed radii. At $R\simeq12$--$13$ kpc, the similar $V_Z$ signals of young and intermediate-age red giants may indicate that the present-day large-scale warp dominates their vertical motions. The difference between the young and intermediate-age profiles at $R\simeq9$--$12$ kpc may reflect a smaller effective kinematic warp onset radius for the intermediate-age populations. This interpretation is consistent with the kinematic results of \citet{2019A&A...627A.150R}, who found warp onset radii of approximately $10.5$--$11$ kpc for red giant branch stars and $12$--$12.5$ kpc for OB stars. \citet{2025A&A...701A.136C} found that the positive $V_Z$ signal appears at approximately $10$ kpc for thin-disk RR Lyrae, compared with approximately $11$ kpc for CCs, also supporting this interpretation. The localized age differences could also arise from residual vertical motions induced by earlier perturbations and superposed on the present-day large-scale warp response, or from pre-existing differences in the populations' vertical structure. In particular, intermediate-age stars may have a more vertically extended distribution than the young stars, consistent with the observed increase in disk scale height with stellar age \citep{2017MNRAS.471.3057M}. Such structural differences could lead to different amplitudes or phases of their response to the same time-dependent perturbation.

\subsection{Robustness Tests}
\label{subsec:robustness}

To assess the robustness of the age-resolved $V_Z(R)$ profiles, we examine the effects of spatial binning, spatial coverage, sample selection, distance scale, and spectroscopic-age uncertainties. To quantify the intermediate-age enhancement, we calculate the $V_Z$ differences relative to the $0$--$2$ Gyr population in each radial bin, and average these differences over $9\leq R<12$ kpc, defined as $\Delta V_Z^{a}\equiv\left\langle V_Z^{a}(R)-V_Z^{0-2}(R)\right\rangle_{9\leq R<12\,\mathrm{kpc}}$, where $a$ denotes the $2$--$4$ or $4$--$6$ Gyr population. Table~\ref{tab:robustness} summarizes the results.

For spatial binning, we vary one setting at a time, testing $\Delta R=0.4$, 0.75, and 1.0 kpc and $\Delta\phi=10^\circ$. We also test shifts of 0.25 kpc in the radial grid and $2.5^\circ$ in the azimuthal grid. We reduce the minimum number of stars per cell from 20 to 15 or 10. We report the ranges of $\Delta V_Z^{a}$ obtained across these settings.

For spatial coverage, we use only $R$--$\phi$ cells that are valid in all three age bins ($0$--$2$, $2$--$4$, and $4$--$6$ Gyr). We retain radial bins containing at least four common cells, leaving 122 cells in 15 radial bins. We calculate $\Delta V_Z^{a}$ using these matched cells and estimate their uncertainties with the bootstrap--Monte Carlo procedure in Section~\ref{subsec:age_profiles}.

For sample selection, we recompute the red-giant profiles using azimuthal cuts of $|\phi|<10^\circ$, $20^\circ$, and $30^\circ$, as well as the two one-sided intervals $-20^\circ<\phi<0^\circ$ and $0^\circ<\phi<20^\circ$. We also vary the vertical selection by adopting $|Z|<1.0$ and $2.0$ kpc and analyzing the $Z>0$ and $Z<0$ subsamples separately. We report the ranges of $\Delta V_Z^{a}$ for the azimuthal and vertical tests separately. To test the effect of RC stars, we repeat the analysis using the non-RC subset.

For the distance scale, we recompute the RC profiles using photometric distances instead of BJ21 geometric distances (Appendix~\ref{app:distance}). The $2$--$4$ and $4$--$6$ Gyr RC profiles lie systematically above the CC profile over $R\simeq9$--$12$ kpc and are broadly consistent with the red-giant profiles at the same ages.

For spectroscopic-age uncertainties, we generate 2000 Monte Carlo realizations by sampling stellar ages according to their uncertainties. In each realization, we reassign the stars to age bins and repeat the subsequent analysis to construct the radial $V_Z$ profiles. For the full red-giant sample, we obtain $\Delta V_Z^{2-4}=1.47^{+0.22}_{-0.23}$ \kms\ and $\Delta V_Z^{4-6}=1.51^{+0.25}_{-0.25}$ \kms, reported as medians with 16th--84th percentile intervals reflecting age perturbations alone. Both $\Delta V_Z$ contrasts remain positive in every realization for the full red-giant and non-RC samples.

Together, these tests support the robustness of the localized $V_Z$ enhancement in the intermediate-age populations.

\begin{deluxetable}{lcc}
\tablecaption{Summary of Robustness Tests
\label{tab:robustness}}
\tablehead{
\colhead{Analysis} &
\colhead{$\Delta V_Z^{2-4}$} &
\colhead{$\Delta V_Z^{4-6}$} \\
\colhead{} & \colhead{(\kms)} & \colhead{(\kms)}
}
\startdata
Baseline profiles           & $1.48\pm0.46$         & $1.49\pm0.52$ \\
Spatial binning             & 1.23--1.49            & 1.14--1.73 \\
Minimum stars per cell      & 0.99--1.48            & 0.95--1.48 \\
Azimuthal selections        & 1.53--1.75            & 1.69--2.07 \\
Vertical selections         & 1.40--1.72            & 1.52--1.83 \\
Non-RC sample               & $1.59\pm0.47$         & $1.66\pm0.58$ \\
Common $R$--$\phi$ coverage  & $1.66\pm0.43$         & $2.05\pm0.45$ \\
Age-error Monte Carlo       & $1.47^{+0.22}_{-0.23}$ & $1.51^{+0.25}_{-0.25}$ \\
\enddata
\tablecomments{
Ranges show the minimum and maximum values obtained with different
settings. 
}
\end{deluxetable}

\section{Summary}
\label{sec:summary}

Using APOGEE DR19 spectroscopy, we construct a spectroscopic-age
catalog of 222,355 red giants. The median relative age uncertainty
is $20.4\%$. Combining this catalog with Gaia astrometry and BJ21
geometric distances, we select 168,824 stars spanning $0$--$10$ Gyr.
We divide this sample into five age bins. We compare the age-resolved $V_Z(R)$ profiles with those of 747 CCs and 2,186 OCs younger than 1 Gyr.

The radial $V_Z$ profiles reveal three main features:

(1) Young red giants ($0$--$2$ Gyr), CCs, and OCs show broadly consistent radial $V_Z$ profiles, indicating that the youngest APOGEE red giants trace the same warp-related vertical motion as established young tracers within the surveyed region.

(2) Over $9\leq R<12$ kpc, the $2$--$4$ and $4$--$6$ Gyr populations in the full red-giant sample have mean $V_Z$ values higher than those of the $0$--$2$ Gyr population by $1.48\pm0.46$ and $1.49\pm0.52$ \kms, respectively.

(3) In the outer disk at $R\simeq12$--$13$ kpc, the $0$--$6$ Gyr red-giant profiles are consistent within their uncertainties.

These results remain robust to variations in spatial binning, spatial coverage, sample selection, distance scale, and spectroscopic-age uncertainties.

The present-day large-scale warp may dominate the outer-disk motions, while the intermediate-age enhancement may reflect a smaller effective onset radius. Possible explanations include the combined effects of earlier perturbations and the present-day warp, or differences in the populations' vertical structure. Broader spatial coverage, more precise ages, and dynamical modeling incorporating the observed stellar spatial distribution will help distinguish these explanations.

\section*{Data Availability}
The machine-readable spectroscopic-age catalog of APOGEE DR19 red giants is available on Zenodo at \url{https://doi.org/10.5281/zenodo.23308202}.

\begin{acknowledgments}
We thank the anonymous referee for the helpful comments. This work was supported by the National Natural Science Foundation of China (NSFC) through the projects 12322306, 12373028, 12633007, 12622304, 12573023, and 12133002. S.~W. and X.~C. acknowledge support from the Youth Innovation Promotion Association of the Chinese Academy of Sciences (CAS; grant Nos.~2023065 and 2022055). Z.~H. acknowledges support from the Fundamental Research Funds of China West Normal University (23kc011, 25KE031) and the Project of the Natural Science Foundation of Sichuan Province (Grant No.~2026YFTX0019). This work has made use of APOGEE DR19 data products from the Sloan Digital Sky Survey V (SDSS-V). Funding for the Sloan Digital Sky Survey V has been provided by the Alfred P.~Sloan Foundation, the Heising-Simons Foundation, the National Science Foundation, and the Participating Institutions. SDSS acknowledges support and resources from the Center for High-Performance Computing at the University of Utah. SDSS telescopes are located at Apache Point Observatory, funded by the Astrophysical Research Consortium and operated by New Mexico State University, and at Las Campanas Observatory, operated by the Carnegie Institution for Science. The SDSS web site is \url{www.sdss.org}. This publication makes use of data products from the Two Micron All Sky Survey, which is a joint project of the University of Massachusetts and the Infrared Processing and Analysis Center/California Institute of Technology, funded by the National Aeronautics and Space Administration and the National Science Foundation. This work has made use of data from the European Space Agency (ESA) mission {\it Gaia} (\url{https://www.cosmos.esa.int/gaia}), processed by the {\it Gaia} Data Processing and Analysis Consortium (DPAC, \url{https://www.cosmos.esa.int/web/gaia/dpac/consortium}). Funding for the DPAC has been provided by national institutions, in particular the institutions participating in the {\it Gaia} Multilateral Agreement.
\end{acknowledgments}

\facilities{Gaia, SDSS, 2MASS}

\appendix
\makeatletter
\@addtoreset{figure}{section}
\makeatother
\renewcommand{\thefigure}{\thesection\arabic{figure}}
\providecommand{\theHfigure}{}
\renewcommand{\theHfigure}{\thesection.\arabic{figure}}
\renewcommand{\theHequation}{\thesection.\arabic{equation}}

\section{Spectroscopic Age Estimation} \label{app:ages}

The age estimates are based on the Astra v0.6.0 \texttt{astraAllStarASPCAP-0.6.0.fits} summary file from the SDSS Science Archive Server, which contains 1,505,140 APOGEE/ASPCAP analysis records before quality cuts and duplicate removal.

\subsection{Asteroseismic Age Calibration Sample} \label{app:seismic_sample}

Cross-matching the APOGEE--Kepler catalog of \citet{miglio2021} with APOGEE DR19 yields 3,314 red giants with asteroseismic ages. Following \citet{2023anders}, we require $0<\tau<15$ Gyr, relative age errors below $30\%$, absolute age errors below 3 Gyr, a signal-to-noise ratio (S/N) of at least 70, ${\rm [Fe/H]}>-1.0$, \texttt{flag\_bad=0} and \texttt{doppler\_flags=0}, and zero abundance-quality flags for Fe and all 15 input elements. Apparently young $\alpha$-rich stars satisfying $[\alpha/{\rm M}]\geq0.14+0.0001(\tau-4.1)^5$, where $\tau$ is in Gyr, are removed. For targets with multiple records, we retain the highest-S/N record, leaving 2,598 calibration stars.

\subsection{Model Training, Calibration, and Validation} \label{app:age_model}

Following \citet{2023anders}, we train an XGBoost regression model using $T_{\rm eff}$, $\log g$, and the $[{\rm X/Fe}]$ ratios for C, N, O, Na, Mg, Al, Si, K, Ca, Ti, V, Mn, Co, Ni, and Ce. The abundance ratios are calculated as $[{\rm X/Fe}]=[{\rm X/H}]-[{\rm Fe/H}]$. $[{\rm Fe/H}]$ is not used as an input feature, reducing the direct dependence of the age estimates on metallicity.

\begin{figure}
\centering
\includegraphics[width=0.60\linewidth]{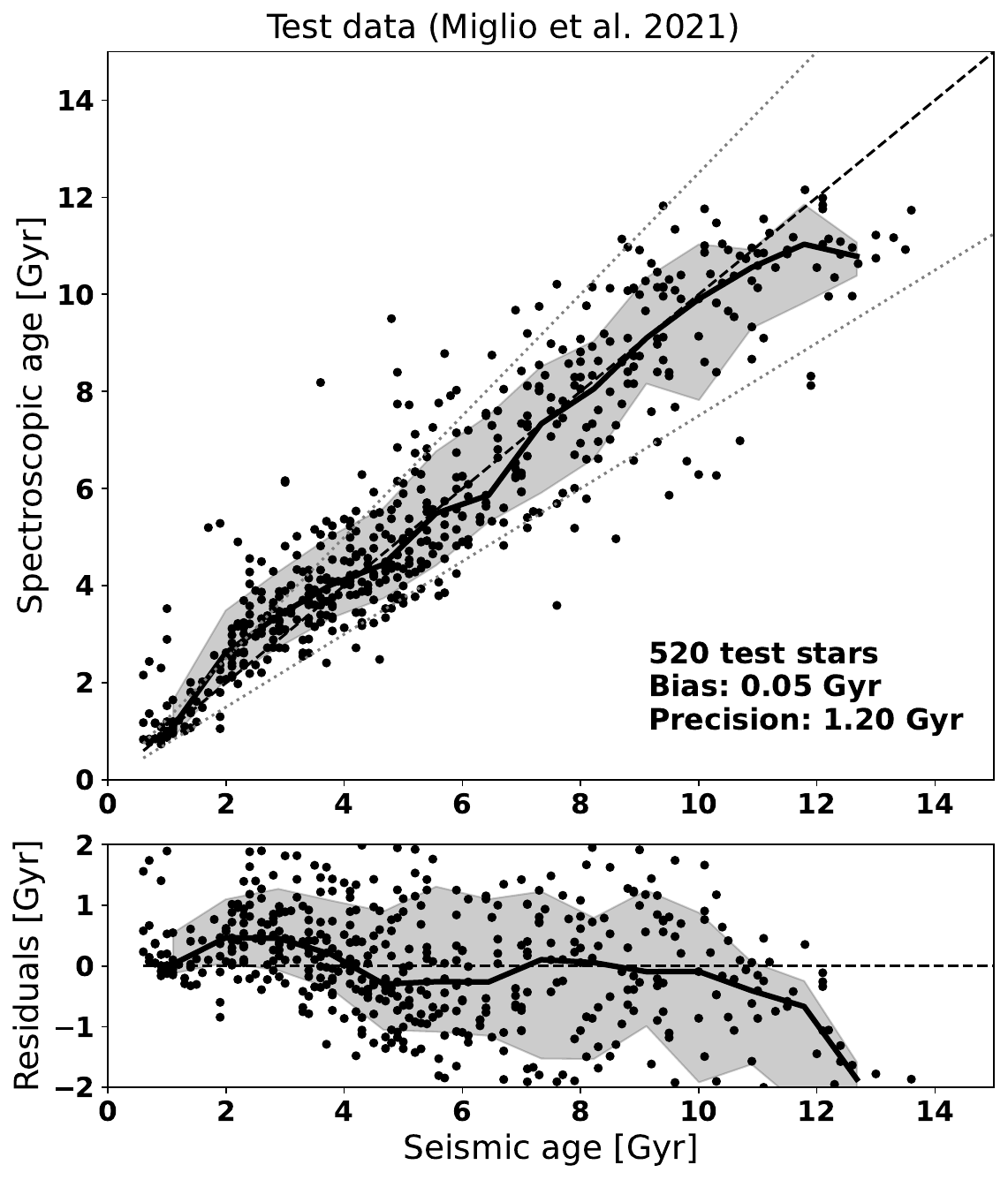}
\caption{Comparison between XGBoost-predicted and asteroseismic ages for the 520-star test set. The upper panel compares the predicted and asteroseismic ages; the dashed line marks the one-to-one relation, and the dotted lines indicate $\pm25\%$ deviations. The lower panel shows the residuals $\tau_{\rm XGB}-\tau_{\rm seis}$, where XGB and seis denote the XGBoost-predicted and asteroseismic ages, respectively; the dashed line marks zero. In both panels, the solid black curves and gray shaded regions indicate the binned medians and 16th--84th percentile ranges.}
\label{fig:age_validation}
\end{figure}

We split the calibration sample into an $80\%$ training set and a $20\%$ test set, with 2,078 and 520 stars, respectively. XGBoost hyperparameters are tuned by three-fold cross-validation and grid search on the training set; the adopted values are \texttt{learning\_rate=0.005}, \texttt{max\_depth=6}, \texttt{min\_child\_weight=5}, \texttt{n\_estimators=1500}, and \texttt{subsample=0.6}. On the test set, the XGBoost predictions give an RMSE of 1.20 Gyr, $R^2=0.848$, and a mean residual of 0.05 Gyr. As shown in Figure~\ref{fig:age_validation}, the model has a small overall bias, slightly overestimating ages at about $2$--$3$ Gyr and showing underestimation trends at $5$--$6$ Gyr and above 10 Gyr. Following \citet{2023anders}, we use the test-set residuals to derive a small a posteriori correction. We fit a fourth-order polynomial to the binned median residuals and estimate age uncertainties with a second-order polynomial fit to the residual dispersion. These individual uncertainty estimates are propagated in the Monte Carlo robustness test described in Section~\ref{subsec:robustness}.

\subsection{DR19 Red Giant Age Catalog} \label{app:age_catalog}

We apply the trained model to APOGEE DR19 red giants with $4300<T_{\rm eff}<5100$ K, $2.2<\log g<3.4$, ${\rm [Fe/H]}>-1.0$, ${\rm S/N}\geq50$, $\sigma_{T_{\rm eff}}<150$ K, complete input features, and valid quality flags. This yields 252,613 age-prediction records. We apply the residual correction from Appendix~\ref{app:age_model} and retain the highest-S/N record for each \texttt{sdss\_id}. The final catalog contains 222,355 stars, with median absolute and relative age uncertainties of 1.11 Gyr and $20.4\%$, respectively.

For each star, the catalog records the SDSS identifier (\texttt{SDSSID}), right ascension (\texttt{ra}), declination (\texttt{dec}), raw XGBoost prediction (\texttt{AgeXGB}), calibrated age (\texttt{AgeXGBCal}), empirical age uncertainty (\texttt{e\_AgeXGB}), and age-quality flags (\texttt{FlagXGB}). We flag entries that lie outside the training-sample parameter range and/or are classified as apparently young $\alpha$-rich stars. We exclude 3,616 stars marked as \texttt{BLUER\_THAN\_TRAINING\_SET}, \texttt{REDDER\_THAN\_TRAINING\_SET}, and/or \texttt{APPARENTLY\_YOUNG\_ALPHA\_RICH}, retaining 218,739 stars for subsequent sample selection. In the released catalog, one non-positive empirical age uncertainty is left blank, and the corresponding star is flagged as \texttt{INVALID\_AGE\_UNCERTAINTY}. This star was retained in the initial kinematic sample but excluded from the $0$--$10$ Gyr analysis because its age is 13.19 Gyr.

\begin{figure}
\centering
\includegraphics[width=0.80\linewidth]{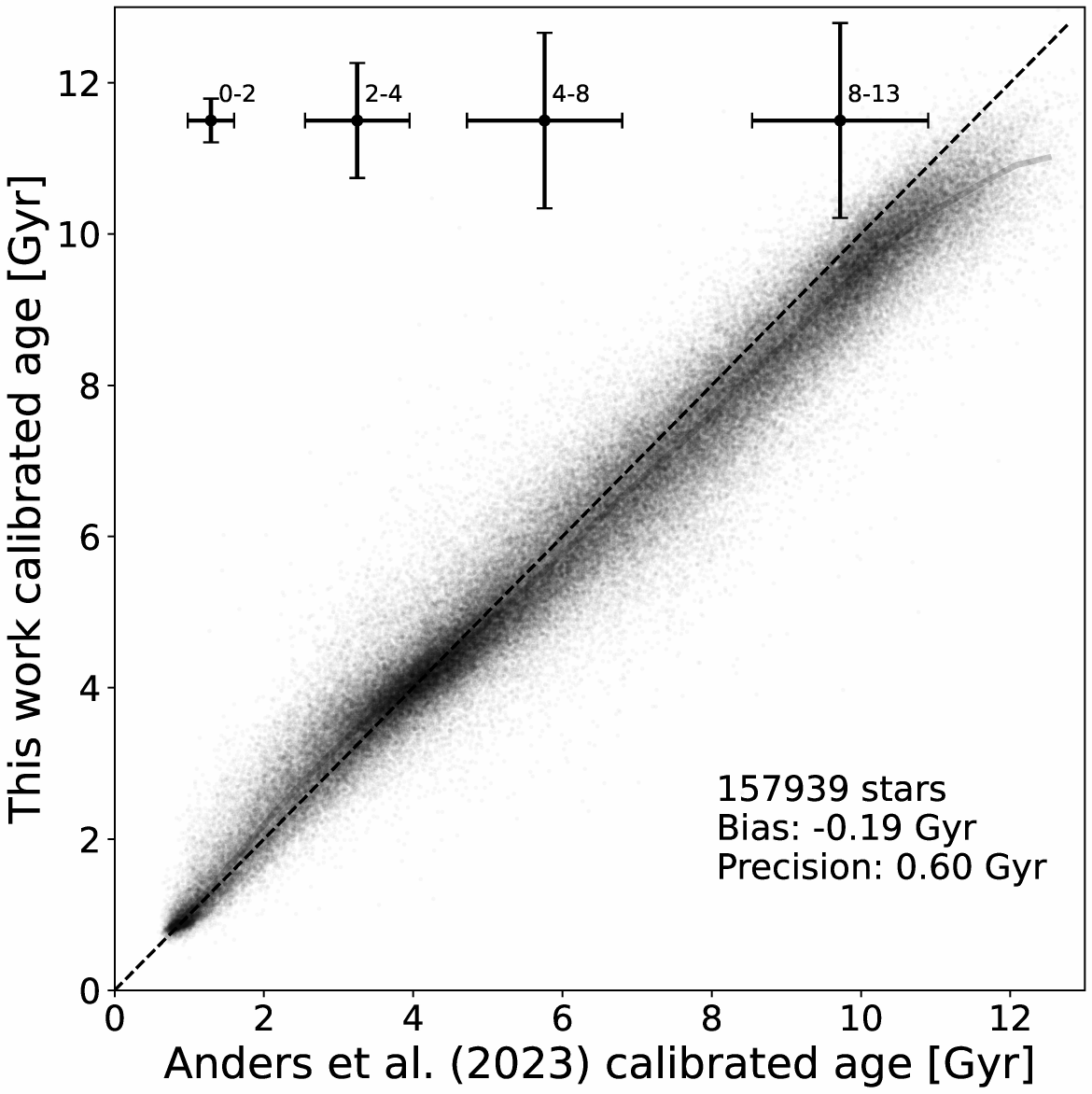}
\caption{Corrected APOGEE Data Release 19 (DR19) ages compared with the corrected Data Release 17 (DR17) ages of \citet{2023anders} for 157,939 matched stars. The dashed line marks the one-to-one relation, and the gray curve shows the binned median. Black points at the top show binned medians; their horizontal and vertical error bars show the median age uncertainties of the DR17 and DR19 catalogs, respectively.}
\label{fig:anders_comparison}
\end{figure}

Figure~\ref{fig:anders_comparison} compares our corrected ages with the corrected APOGEE Data Release 17 (DR17) ages of \citet{2023anders} for 157,939 matched stars. The mean difference, defined as $\tau_{\rm DR19}-\tau_{\rm DR17}$, is $-0.19$ Gyr, with a standard deviation of 0.60 Gyr. The two catalogs agree broadly, although the DR19 ages are slightly lower at the oldest ages. The two catalogs were produced using similar methods and overlapping asteroseismic calibration samples. This comparison tests consistency between APOGEE data releases rather than providing an independent validation.

\section{Photometric Distances and Vertical-velocity Profiles of RC Stars}
\label{app:distance}

To test the sensitivity of the age-resolved $V_Z(R)$ profiles to the
adopted distance scale, we derive independent photometric distances
for the RC subset. We use the 2MASS $K_s$ photometry included in the APOGEE DR19 catalog \citep{2006AJ....131.1163S}. We calibrate the extinction and $K_s$-band RC
absolute magnitude using the APOGEE--Gaia sample, and then recompute
the Galactocentric phase-space coordinates and $V_Z(R)$ profiles of
the RC stars.

\subsection{Photometric Distance Calibration}
\label{app:photometric_distance}

To calibrate the intrinsic color of RC stars, we select RC candidates from the parent red-giant age catalog using the criteria defined in Section~\ref{subsec:rg_sample}. We require ${\rm S/N}\geq70$, $\sigma_{\log g}<0.08$ dex, $\sigma_{[\mathrm{M}/\mathrm{H}]}<0.02$ dex, $\sigma_{T_{\rm eff}}<150$ K, ${\rm RUWE}<1.4$, and valid Gaia $G_{\rm BP}$ and $G_{\rm RP}$ photometry. These criteria yield a high-quality calibration sample of 54,785 RC stars.

We calibrate $(G_{\rm BP}-G_{\rm RP})_0$ using the blue-edge method \citep{2014ApJ...788L..12W,2017AJ....153....5J,2019ApJ...877..116W,zhao2024}. We divide the calibration sample into
bins of 80 K in $T_{\rm eff}$ and 0.2 dex in $[\mathrm{M}/\mathrm{H}]$. For each bin containing at least 30 stars, we select the bluest 5\% and adopt their median color as the blue edge. Fitting the resulting 32 calibration points gives $(G_{\rm BP}-G_{\rm RP})_0=-19.2356(\log_{10}T_{\rm eff})^2+137.7374\log_{10}T_{\rm eff}+0.0807{\rm [M/H]}^2+0.1540{\rm [M/H]}+0.0446\log g-245.2871$, where $T_{\rm eff}$ is in K. The fit residuals have a root-mean-square error of 0.011 mag. 

We apply this relation to RC candidates with valid Gaia photometry and calculate the color excess as $E(G_{\rm BP}-G_{\rm RP})=(G_{\rm BP}-G_{\rm RP})_{\rm obs}-(G_{\rm BP}-G_{\rm RP})_0$. We then adopt the extinction relations of \citet{2019ApJ...877..116W}, $A_{K_s}=0.186E(G_{\rm BP}-G_{\rm RP})$ and $A_V=2.394E(G_{\rm BP}-G_{\rm RP})$.

We calibrate the RC $K_s$-band absolute magnitude using 3,234 low-extinction stars with high-quality Gaia parallaxes corrected for the parallax zero point \citep{2021lindegren}. We require $A_V<0.1$ mag, ${\rm RUWE}<1.4$, $\varpi_{\rm corr}>1$ mas, and $\sigma_\varpi/\varpi_{\rm corr}<0.05$, where $\varpi_{\rm corr}$ and $\sigma_\varpi$ are the corrected parallax and
its uncertainty, respectively. The absolute magnitude is calculated as $M_{K_s}=K_s-A_{K_s}-5\log_{10}(d_\varpi/{\rm kpc})-10$, where $d_\varpi(\mathrm{kpc})=1/\varpi_{\rm corr}(\mathrm{mas})$. Robust $3\sigma$ clipping leaves 3,027 stars, with a median $M_{K_s}=-1.582$ mag and a standard deviation of 0.106 mag. We adopt the sample median, $M_{K_s}=-1.58$ mag, as the reference absolute magnitude for the photometric-distance calculation.

We calculate the RC photometric distance as $d_{\rm RC}({\rm kpc})=10^{(K_s-A_{K_s}-M_{K_s}-10)/5}$. The 0.106 mag dispersion in $M_{K_s}$ corresponds to a characteristic fractional distance scatter of approximately $5\%$ from the absolute-magnitude calibration alone. We further require $G,G_{\rm BP},G_{\rm RP}\leq18$ mag, $\sigma_{G_{\rm BP}},\sigma_{G_{\rm RP}}\leq0.01$ mag, $6\leq K_s\leq14$ mag, and $\sigma_{K_s}\leq0.05$ mag. Applying these photometric-quality cuts to the full RC candidate sample leaves 59,185 stars with photometric distances.

\subsection{Vertical-velocity Profiles}
\label{app:rc_photometric_profiles}

Using the photometric distances derived above, we recompute the Galactocentric positions and velocities of the RC stars. We apply the same phase-space selection, age binning, and radial-profile estimator as in the fiducial analysis (Sections~\ref{subsec:rg_sample}--\ref{subsec:age_profiles}), obtaining 46,699 RC stars within $|\phi|<40^\circ$, of which 45,985 fall in the $0$--$10$ Gyr age bins.

\begin{figure*}
\centering
\includegraphics[width=\textwidth]
{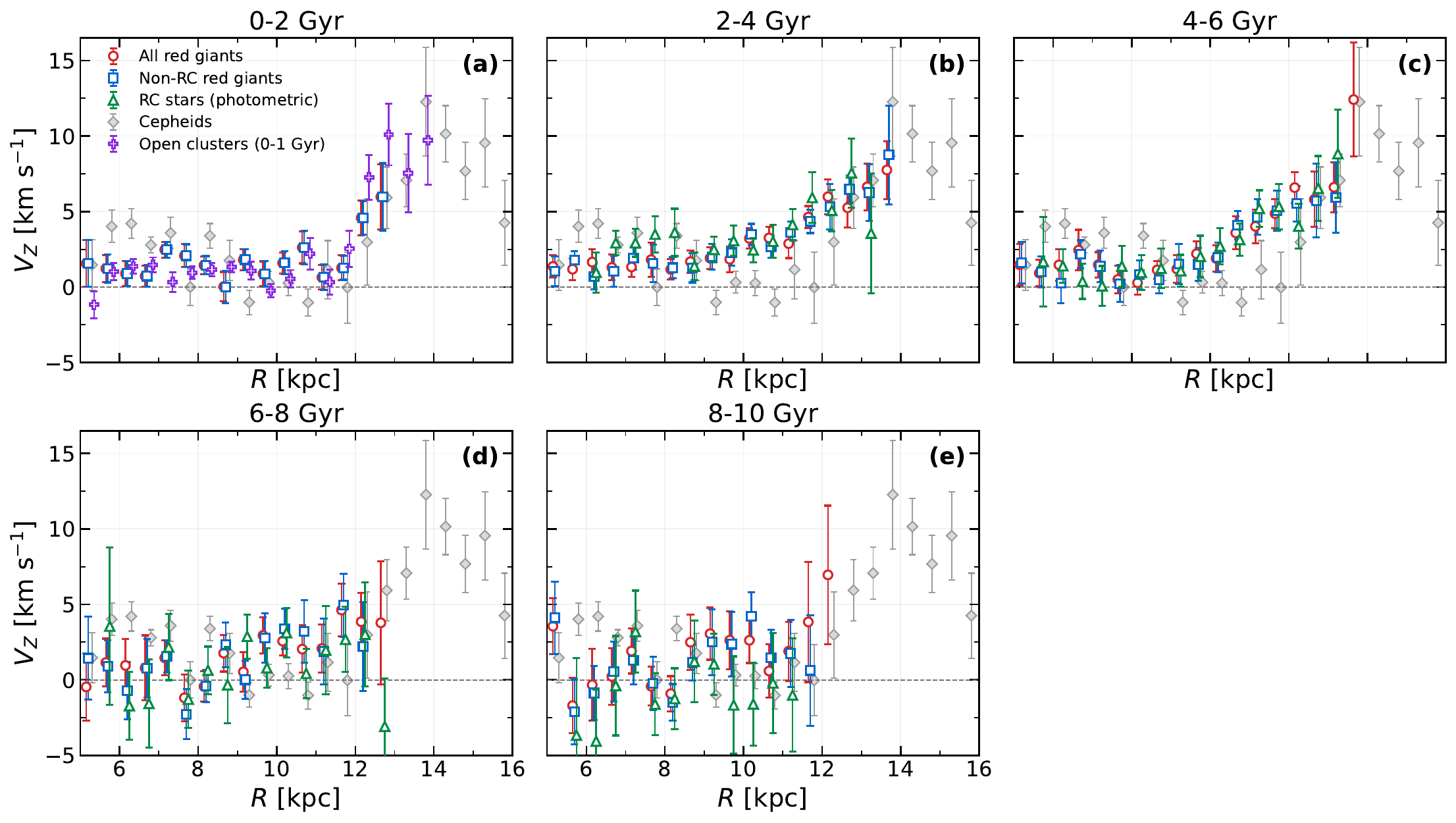}
\caption{Age-resolved $V_Z(R)$ profiles for RC stars with photometric distances. Symbols follow Figure~\ref{fig:multitracer_vz}, with green triangles indicating the RC profiles.}
\label{fig:rc_photometric_profiles}
\end{figure*}

Figure~\ref{fig:rc_photometric_profiles} compares the resulting RC profiles with the full red-giant and non-RC profiles based on BJ21 geometric distances. The $0$--$2$ Gyr RC sample is too sparse to construct a radial profile. For the $2$--$4$ and $4$--$6$ Gyr populations, the RC profiles are broadly consistent with the corresponding full red-giant profiles and retain systematically higher $V_Z$ than the CC population over $R\simeq9$--$12$ kpc. The main age-dependent $V_Z$ feature persists when photometric distances are adopted for the RC stars.

\clearpage
\bibliographystyle{aasjournalv7.1}
\bibliography{refs}

\end{document}